\documentclass[twocolumn,epjc3]{svjour3}

\RequirePackage{graphicx}

\RequirePackage{amsmath,amsfonts}
\RequirePackage[numbers,sort&compress]{natbib}
\RequirePackage[colorlinks,citecolor=blue,urlcolor=blue,linkcolor=blue]{hyperref}

\journalname{Eur. Phys. J. C}

\begin{document}

\title{AIC, BIC, Bayesian evidence against the interacting dark energy model}

\author{Marek Szyd{\l}owski\thanksref{oauj,csrc,e-ms}
\and
Adam Krawiec\thanksref{ieuj,csrc,e-ak}
\and
Aleksandra Kurek\thanksref{oauj,e-ok}
\and
Micha{\l} Kamionka\thanksref{aiuw,e-mk}
}

\thankstext{e-ms}{marek.szydlowski@uj.edu.pl}
\thankstext{e-ak}{adam.krawiec@uj.edu.pl}
\thankstext{e-ok}{alex@oa.uj.edu.pl}
\thankstext{e-mk}{kamionka@astro.uni.wroc.pl}

\institute{Astronomical Observatory,
Jagiellonian University, Orla 171, 30-244 Krak{\'o}w, Poland \label{oauj}
\and
Institute of Economics and Management,
Jagiellonian University, {\L}ojasiewicza 4, 30-348 Krak{\'o}w, Poland \label{ieuj}
\and
Mark Kac Complex Systems Research Centre, Jagiellonian University,
Reymonta 4, 30-059 Krak{\'o}w, Poland \label{csrc}
\and
Astronomical Institute, University of Wroc{\l}aw,
ul. Kopernika 11, 51-622 Wroc{\l}aw, Poland \label{aiuw}
}

\date{Received: date / Accepted: date}

\maketitle

\begin{abstract}
Recent astronomical observations have indicated that the Universe is in the phase of accelerated expansion. While there are many cosmological models which try to explain this phenomenon, we focus on the interacting $\Lambda$CDM model where the interaction between the dark energy and dark matter sectors takes place. This model is compared to its simpler alternative---the $\Lambda$CDM model. To choose between these models the likelihood ratio test was applied as well as the model comparison methods (employing Occam's principle): the Akaike information criterion (AIC), the Bayesian information criterion (BIC) and the Bayesian evidence. Using the current astronomical data: SNIa (Union2.1), $h(z)$, BAO, Alcock--Paczynski test and CMB we evaluated both models. The analyses based on the AIC indicated that there is less support for the interacting $\Lambda$CDM model when compared to the $\Lambda$CDM model, while those based on the BIC indicated that there is the strong evidence against it in favor the $\Lambda$CDM model. Given the weak or almost none support for the interacting $\Lambda$CDM model and bearing in mind Occam's razor we are inclined to reject this model.
\end{abstract}

\section{Introduction}

Recent observations of type Ia supernova (SNIa) provide the main evidence that current Universe is in an accelerating phase of expansion \cite{Riess:1998cb}. Cosmic microwave background (CMB) data indicate that the present Universe has also a negligible space curvature \cite{Spergel:2006hy}. Therefore if we assume the Friedmann-Robertson-Walker (FRW) model in which effects of nonhomogeneities are neglected, then the acceleration must be driven by a dark energy component $X$ (matter fluid violating the strong energy condition $\rho_{\text{X}}+3p_{\text{X}}\geq 0)$. This kind of energy represents roughly $70\%$ of the matter content of the current Universe. Because the nature as well as mechanism of the cosmological origin of the dark energy component are unknown some alternative theories try to eliminate the dark energy by modifying the theory of gravity itself. The main prototype of this kind of models is a class of covariant brane models based on the Dvali-Gabadadze-Porrati (DGP) model \cite{Dvali:2000hr} as generalized to cosmology by Deffayet \cite{Deffayet:2000uy}. The simplest explanation of a dark energy component is the cosmological constant with effective equation of state $p=-\rho$ but appears the problem of its smallness and hence its relatively recent dominance. Although the $\Lambda$CDM model offers possibility of explanation of observational data it is only effective theory which contain the enigmatic theoretical term---the cosmological constant $\Lambda$. Other numerous candidates for dark energy description have also been proposed like to evolving scalar field \cite{Peebles:1987ek} usually referred as quintessence, the phantom energy \cite{Caldwell:1999ew, Dabrowski:2003jm}, the Chaplygin gas \cite{Kamenshchik:2001cp} etc. Some authors believed that the dark energy problem belongs to the quantum gravity domain \cite{Witten:2000zk}.

Recent Planck observations still favor the Standard Cosmological Model \cite{Ade:2013zuv}, especially for the high multipoles. However in this model there are some problems with understanding the values of density parameters for both dark matter and dark energy. The question is: why energies of vacuum energy and dark matter are of the same order for current Universe? The very popular methodology to solve this problem is to treat coefficient equation of state as a free parameter, i.e. the wCDM model which should be estimated from the astronomical and astrophysical data. The observations from CMB and baryon acoustic oscillation (BAO) data sets give $w_x=-1.13^{+0.24}_{-0.23}$ with $95\%$ confidence levels \cite{Ade:2013zuv}.

Alternative to this idea of the phantom dark energy mechanism of alleviate the coincidence problem is to consider the interaction between dark matter and dark energy, interacting model. Many authors investigated observational constraints of the interacting model. A. Costa et al. \cite{Costa:2013sva} concluded that the interacting models becomes in agreement with the admissible observational data which can provide some argument towards consistency of measured density parameters. W. Yang et al. \cite{Yang:2014gza} constrained some interacting models under the choice of ansatz for transfer energy mechanism. From this investigation the joining geometrical tests show a stricter constraint on the interacting model if we included the information from large scale structure ($f\sigma_{8}(z)$ data) of the Universe. These authors have found the interaction rate in the $3\sigma$ region. This means that the recent cosmic observations favor but rather the small interaction between the both dark sectors. However, the measurement of redshift-space distortion could rule out a large interaction rate in the $1\sigma$ region. M.~J. Zhang and W.~B. Liu \cite{Zhang:2013zyn} using the type Ia supernova observations, $H(z)$ data (OHD), cosmic microwave background (CMB) and secular Sandage-Loeb (SL) obtained the small value of the interacting parameter: $\delta=-0.019\pm 0.01 (1 \sigma), \pm 0.02 (2\sigma)$.

In all interacting models the specific ansatz for a model of interaction is postulated. There are infinite of such models with a different form of interaction and this is some kind of a theoretical bias or degeneracy, coming from the choice of a potential form in a scalar field cosmology. Szydlowski proposed the idea of estimation of an interaction parameter without any ansatz for the model of interaction \cite{Szydlowski:2005ph}.

These theoretical models are consistent with the observations, they are able to explain the phenomenon of the accelerated expansion of the Universe. But should we really prefer such models over the $\Lambda$CDM one? All observational constraints show that the $\Lambda$CDM model still shows a good fit to the observational data. But from these constraints the small value of interaction is still admissible. To answer this question we should use some model comparison methods to confront existing cosmological models having observations at hand. We choose the information and Bayesian criteria of model selection which are based on Occam's razor (principle), the well-known and effective instrument in science to obtain a definite answer whether the interacting $\Lambda$CDM model can be rejected.

Let us assume that we have $N$ pairs of measurements $(y_i,x_i)$ and that we want to find the relation between the $y$ and $x$ variables. Suppose that we can postulate $k$ possible relations $y\equiv f_i(x,\bar{\theta})$, where $\bar{\theta}$ is the vector of unknown model parameters and $i=1,\dots,k$. With the assumption that our observations come with uncorrelated Gaussian errors with a mean $\mu_i=0$ and a standard deviation $\sigma_i$ the goodness of fit for the theoretical model is measured by the $\chi^2$ quantity given by
\begin{equation}
\chi^2=\sum_{i=1}^{N} \frac{(f_l(x_i,\bar{\theta}) - y_i)^2}{2\sigma_i^2}=-2\ln L,
\end{equation}
where $L$ is the likelihood function. For the particular family of models $f_l$ the best one minimize the $\chi^2$ quantity, which we denote $f_l(x,\hat{\bar{\theta}})$. The best model from our set of $k$ models ${f_1(x,\hat{\bar{\theta}}),\dots,f_k(x,\hat{\bar{\theta}})}$ could be the one with the smallest value of $\chi^2$ quantity. But this method could give us misleading results. Generally speaking for more complex model the value of $\chi^2$ is smaller, thus the most complex one will be choose as the best from our set under consideration.

\begin{figure}[h]
\centering
\includegraphics[scale=0.5]{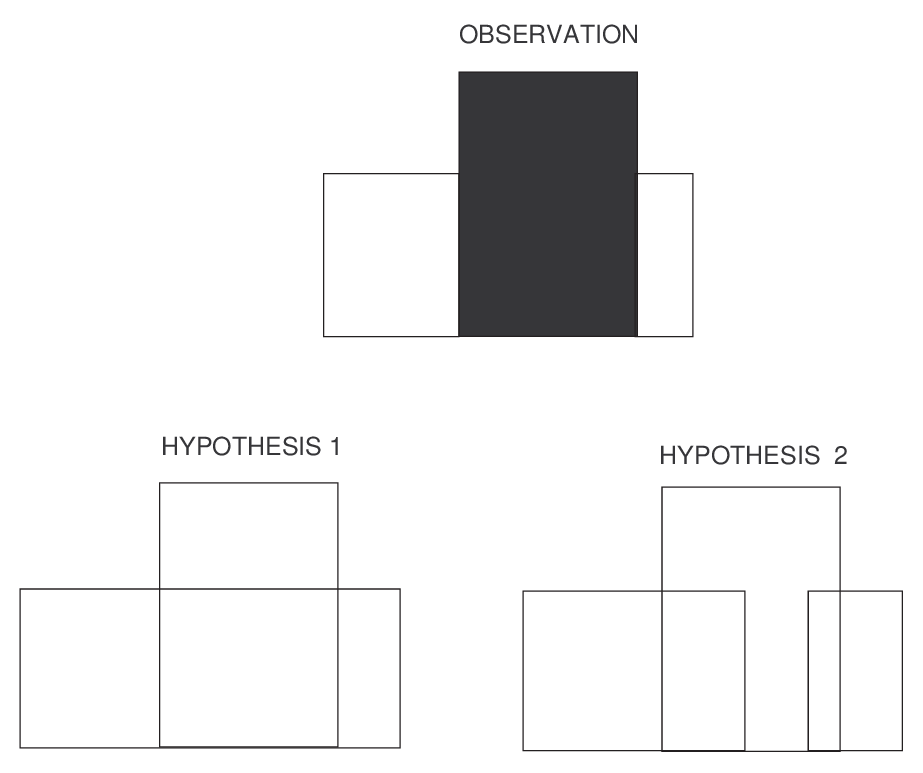}
\caption{The illustration of Occam's principle.}
\label{Fig:1}
\end{figure}

A clue is given by Occam's principle known also as Occam's razor: ``If two models describe the observations equally well, choose the simplest one.'' This principle has an aesthetic as well as empirical justification. Let us quote the simple example which illustrates this rule \cite{MacKay:2003it}. In Figure \ref{Fig:1} it is observed the black box and the white one behind it. One can postulate two models: first, there is one box behind the black box, second, there are two boxes of identical height and color behind the black box. Both models explain our observations equally well. According to Occam's principle we should accept the explanation which is simpler so that there is only one white box behind the black one. Is not it more probable that there is only one box than two boxes with the same height and color?

We could not use this principle directly because the situations when two models explain the observations equally well are rare. But in the information theory as well as in the Bayesian theory there are methods for model comparison which include such a rule.

In the information theory there are no true models. There is only reality which can be approximated by models, which depend on some number of parameters. The best one from the set under consideration should be the best approximation to the truth. The information lost when truth is approximated by model under consideration is measured by the so called Kullback-Leibler (KL) information so the best one should minimize this quantity. It is impossible to compute the KL information directly because it depends on truth which is unknown. Akaike \cite{Akaike:1974nl} found an approximation to the KL quantity which is called the Akaike information criterion (AIC) and is given by
\begin{equation}
\text{AIC}=-2\ln \mathcal{L} +2d,
\end{equation}
where $\mathcal{L}$ is the maximum of the likelihood function and $d$ is the number of model parameters. A model which is the best approximation to the truth from a set of models under consideration has the smallest value of the AIC quantity. It is convenient to evaluate the differences between the AIC quantities computed for the rest of models from the set and the AIC for the best one. Those differences ($\Delta_{\text{AIC}}$) are easy to interpret and allow a quick `strength of evidence' for a considered model with respect to the best one. The models with $0 \le \Delta_{\text{AIC}}\le 2$ have substantial support (evidence), those where $4<\Delta_{\text{AIC}}\le 7$ have considerably less support, while models having $\Delta_{\text{AIC}} > 10 $ have essentially no support with respect to the best model.

It is worth noting that the complexity of the model is interpreted here as the number of its free parameters that can be adjusted to fit the model to the observations. If models under consideration fit the data equally well according to the Akaike rule the best one is with the smallest number of model parameters (the simplest one in such an approach).

In the Bayesian framework the best model (from the model set under consideration) is that which has the largest value of probability in the light of data (so called posterior probability) \cite{Jeffreys:1961}
\begin{equation}
P(M_{i}|D)=\frac{P(D|M_{i})P(M_{i})}{P(D)},
\end{equation}
where $P(M_{i})$ is a prior probability for the model $M_{i}$, $D$ denotes data, $P(D)$ is the normalization constant
\begin{equation}
P(D)= \sum _{i=1}^{k} P(D|M_{i})P(M_{i}).
\end{equation}
And $P(D|M_{i})$ is the marginal likelihood, also called evidence
\begin{equation}
P(D|M_{i})=\int P(D|\bar{\theta},M_{i})P(\bar{\theta}|M_{i}) \ d \bar{\theta} \equiv E_{i},
\end{equation}
where $P(D|\bar{\theta},M_{i})$ is likelihood under model $i$, $P(\bar{\theta}|M_{i})$ is prior probability for ${\bar{\theta}}$ under model $i$.

Let us note that we can include Occam's principle by assuming the greater prior probability for simpler model, but this is not necessary and rarely used in practice. Usually one assume that there is no evidence to favor one model over another which cause to equal value of prior for all models under consideration. It is convenient to evaluate the posterior ratio for models under consideration which in the case with flat prior for models is reduced to the evidence ratio called the Bayes factor
\begin{equation}
B_{ij} = \frac{P(D|M_i)}{P(D|M_j)}.
\end{equation}
The interpretation of twice the natural logarithm of the Bayes factor is as follow: $0<2\ln B_{ij}\leq 2$ as a weak evidence, $2<2\ln B_{ij}\leq 6$ as a positive evidence, $6<2\ln B_{ij}\leq 10$ as a strong evidence and $2\ln B_{ij}> 10$ as a very strong evidence against model $j$ comparing to model $i$. This quantity is our Occam's razor. Let us simplify the problem to illustrate how this principle works here \cite{MacKay:2003it,Trotta:2005ar}.

Assume that $\bar{P}(\bar{\theta}|D,M)$ is the non normalized posterior probability for the vector $\bar{\theta}$ of model parameters. In this notation $E=\int\bar{P}(\bar{\theta}|D,M)d\bar{\theta}$. Suppose that posterior has a strong peak in the maximum: $\bar{\theta}_{\text{MOD}}$. It is reasonable to approximate the logarithm of the posterior by its Taylor expansion in the neighborhood of $\bar{\theta}_{\text{MOD}}$ so we finished with the expression
\begin{align}
\bar{P}(\bar{\theta}|D,M) & = \bar{P}(\bar{\theta}_\text{MOD}|D,M) \times \nonumber \\
& \times \exp \left[-(\bar{\theta}-\bar{\theta}_{\text{MOD}})^T C^{-1}(\bar{\theta}-\bar{\theta}_{\text{MOD}})\right],
\end{align}
where $\left[ C^{-1} \right]_{ij} = -\left[\frac{\partial^2\ln\bar{P}(\bar{\theta}|D,M)}{\partial\theta_i\partial\theta_j}\right]_{\bar{\theta}=\bar{\theta}_\text{MOD}}$. The posterior is approximated by the Gaussian distribution with the mean $\bar{\theta}_\text{MOD}$ and the covariance matrix $C$. The evidence then has a form
\begin{eqnarray}
E & = & \bar{P}(\bar{\theta}_{\text{MOD}}|D,M) \times \nonumber \\
& \times & \int \exp \left[-(\bar{\theta}-\bar{\theta}_{\text{MOD}})^T C^{-1}(\bar{\theta}-\bar{\theta}_{\text{MOD}})\right] \ d \bar{\theta}.
\end{eqnarray}
Because the posterior has a strong peak near the maximum, the most contribution to the integral comes from the neighborhood close to $\bar{\theta}_\text{MOD}$. Contribution from the other region of $\bar{\theta}$ can be ignored, so we can expand the limit of the integral to whole $R^d$. With this assumption one can obtain $E=(2\pi)^{\frac{d}{2}}\sqrt{\det C}\bar{P}(\bar{\theta}_{\text{MOD}}|D,M)= (2\pi)^{\frac{d}{2}}\sqrt{\det C}P(D|\bar{\theta}_\text{MOD},M)P(\bar{\theta}_\text{MOD}|M)$.
Suppose that the likelihood function has a sharp peak in $\hat{\bar{\theta}}$ and the prior for $\bar{\theta}$ is nearly flat in the neighborhood of $\hat{\bar{\theta}}$. In this case $\hat{\bar{\theta}}=\bar{\theta}_{\text{MOD}}$ and the expression for the evidence takes the form $E=\mathcal{L}(2\pi)^{\frac{d}{2}}\sqrt{\det\text{C}}P(\hat{\bar{\theta}}|M)$. The quantity $(2\pi)^{\frac{d}{2}}\sqrt{\det\text{C}}P(\hat{\bar{\theta}}|M)$ is called the Occam factor (OF). When we consider the case with one model parameter with a flat prior $P(\theta|M)=\frac{1}{\Delta\theta}$ the Occam factor OF$=\frac{2\pi\sigma}{\Delta\theta}$ which can be interpreted as the ratio of the volume occupied by the posterior to the volume occupied by prior in the parameter space. The more parameter space wasted by the prior the smaller value of the evidence. It is worth noting that the evidence does not penalize parameters which are unconstrained by the data \cite{Liddle:2006kn}.

As the evidence is hard to evaluate an approximation to this quantity was proposed by Schwarz \cite{Schwarz:1978ed} so called Bayesian information criterion (BIC) and is given by
\begin{equation}
\text{BIC}=-2\ln\mathcal{L}+2d\ln N,
\end{equation}
where $N$ is the number of the data points. The best model from a set under consideration is this which minimizes the BIC quantity. One can notice the similarity between the AIC and BIC quantities though they come from different approaches to model selection problem. The dissimilarity is seen in the so called penalty term: $ad$, which penalize more complex models (complexity is identified here as the number of free model parameters). One can evaluated the factor by which the additional parameter must improve the goodness of fit to be included in the model. This factor must be greater than $a$ so equal to $2$ in the AIC case and equal to $\ln N$ in the BIC case. Notice that the latter depends on the number of the data points.

It can be shown that there is the simple relation between the BIC and the Bayes factor
\begin{equation}
    2 \ln B_{ij} = -(\text{BIC}_i - \text{BIC}_j).
\end{equation}
The quantity $B_{ij}$ is the Bayes factor for the hypothesis (model) $i$ against the hypothesis (model) $j$. We categorize this evidence against the model $j$ taking the following ranking. The evidence against the model $j$ is not worth than bare mention when twice the natural logarithm of the Bayes factor (or minus the difference between BICs) is $0< 2\ln B_{ij} \leq 2$, is positive when $2< 2\ln B_{ij} \leq 6$, is strong when $6< 2\ln B_{ij}\leq 10$ and is very strong when $ 2\ln B_{ij} > 10$.

It should be pointed out that presented model selection methods are widely used in context of cosmological model comparison \cite{Hobson:2002de, Beltran:2005xd, Mukherjee:2005wg, Mukherjee:2005tr, Trotta:2005ar, Niarchou:2003hz, Liddle:2004nh, Saini:2003wq, John:2002gg, Parkinson:2004yx, John:2005bz, Serra:2007id, Biesiada:2007um, Liddle:2006kn, Godlowski:2005tw, Szydlowski:2006ay, Szydlowski:2005xv, Kunz:2006mc, Parkinson:2006ku, Trotta:2007hy, Liddle:2007fy, Kurek:2007tb}. We should keep in mind that conclusions based on such quantities depend on the data at hand. Let us mention again the example with the black box. Suppose that we made a few steps toward this box that we can see the difference between the height of the left and right side of the white box. Our conclusion changes now.

Let us quote the example taken from \cite{John:2005bz}. Assume that we want to compare the Newtonian and Einsteinian theories in the light of the data coming from a laboratory experiment where general relativistic effects are negligible. In this situation the Bayes factor between Newtonian and Einsteinian theories will be close to unity. Whereas comparing the general relativistic and Newtonian explanations of the deflection of a light ray that just grazes the Sun's surface give the Bayes factor $\sim 10^{10}$ in the favor of the first one (and even greater with more accurate data).

We share George Efstathiou's opinion \cite{Efstathiou:2007gz,Chongchitnan:2007eb,Szydlowski:2004jv} that there is no sound theoretical basis for considering the dynamical dark energy, where as we are beginning to see an explanation for a small cosmological constant emerging from more fundamental theory. In our opinion the $\Lambda$CDM model has the status of the satisfactory effective theory. Efstathiou argued why the cosmological constant should be given higher weight as a candidate for dark energy description than dynamical dark energy. In this argumentation Occam's principle is used to point out a more economical model explaining the observational data.

The main aim of this paper is to compare the simplest cosmological model---the $\Lambda$CDM model---with its generalization where the interaction between dark energy and matter sectors is allowed using methods described above.

\section{Interacting $\Lambda$CDM model}

The interacting interpretation of the continuity condition (conservation condition) was investigated in the context of the coincidence problem since
the paper Zimdahl \cite{Zimdahl:2005bk}, for recent developments in this area see Olivares et al. \cite{Olivares:2006jr,Olivares:2007rt}, see also Le
Delliou et al. \cite{LeDelliou:2007am} for discussion recent observational constraints.

Let us consider two basic equations which determine the evolution of FRW cosmological models
\begin{align}\label{eq:1a}
\frac{\ddot{a}}{a}=-\frac{1}{6}(\rho+3p) \\
\label{eq:1b}
\dot{\rho}=-3H(\rho+p).
\end{align}
Equation (\ref{eq:1a}) is called the accelerated equation and equation (\ref{eq:1b}) is the conservation (or adiabatic) condition. Equation (\ref{eq:1a}) can be rewritten to the form analogous to the Newtonian equation of motion
\begin{equation}\label{eq:2}
\ddot{a}=-\frac{\partial V}{\partial a},
\end{equation}
where $V=V(a)$ is potential function of the scale factor $a$. To evaluate $V(a)$ from (\ref{eq:2}) via integration by parts it is useful to rewrite (\ref{eq:1b}) to the new equivalent form
\begin{equation}\label{eq:3}
\frac{d}{dt}(\rho a^3) + p \frac{d}{dt}(a^3)=0.
\end{equation}
From (\ref{eq:1a}) we obtain
\begin{equation}\label{eq:4}
\frac{\partial V}{\partial a}=\frac{1}{12}(\rho+3p)d(a^2).
\end{equation}
It is convenient to calculate pressure $p$ from (\ref{eq:3}) and then substitute to (\ref{eq:4}). After simple calculations we obtain from (\ref{eq:4})
\begin{equation}\label{eq:5}
\frac{\partial V}{\partial a} = - \frac{1}{6}\left[a^2 \frac{d\rho}{da}+\rho d(a^2) \right].
\end{equation}
Therefore
\begin{equation}\label{eq:6}
V(a)=-\frac{\rho a^2}{6}.
\end{equation}
In formula (\ref{eq:6}) $\rho$ means the effective energy density of the fluid filling the Universe.

We find the very simple interpretation of (\ref{eq:1a}): the evolution of the Universe is equivalent to the motion of the particle of unit mass in the potential well parameterized by the scale factor. In the procedure of reduction of the problem of FRW evolution to the problem of investigation dynamical system of a Newtonian type we only assume that the effective energy density satisfies the conservation condition. We do not assume the conservation condition for each energy component (or non-interacting matter sectors).

Equations (\ref{eq:1a}) and (\ref{eq:1b}) admit the first integral which is usually called the Friedmann first integral. This first integral has a simple interpretation in the particle-like description of the FRW cosmology, namely energy conservation
\begin{equation}\label{eq:7}
\frac{\dot{a}^2}{2}+V(a)=E=-\frac{k}{2},
\end{equation}
where $k$ is the curvature constant and $V$ is given by formula (\ref{eq:6}).

Let us consider the universe filled with the two components fluid
\begin{equation}\label{eq:8}
\rho=\rho_{\text{m}} + \rho_X, \quad p=0+w_X\rho_X,
\end{equation}
where $\rho_{\text{m}}$ means energy density of usual dust matter and $\rho_X$ denotes energy density of dark energy satisfying the equation of state $p_X=w_X\rho_X$, where $w_X=w_X(a)$. Then equation (\ref{eq:3}) can be separated on the dark matter and dark energy sectors which in general can interact
\begin{align}\label{eq:9}
& \frac{d}{dt}(\rho_{\text{m}} a^3) + 0 \cdot \frac{d}{dt}(a^3)=\Gamma \\
& \frac{d}{dt}(\rho_X a^3) + w_X(a)\rho_X \frac{d}{dt}(a^3)=-\Gamma
\end{align}
In our previous paper \cite{Szydlowski:2005kv} it was assumed that
\begin{equation}\label{eq:10}
\Gamma=\alpha a^n \frac{\dot{a}}{a},
\end{equation}
which able us to integrate (\ref{eq:9}) which gives
\begin{equation}\label{eq:11}
\rho_m= \frac{C}{a^3}+\frac{\alpha}{n}a^{n-3}
\end{equation}
\begin{equation}\label{eq:12}
\frac{d \rho_X}{da}+\frac{3}{a}(1+w_X(a))\rho_X=-\alpha a^{n-4}.
\end{equation}
The solution of homogeneous equation (\ref{eq:12}) can be written in terms of average $\overline{w_X}(a)$ as
\begin{equation}\label{eq:13}
\rho_X=\rho_{X,0}a^{-3(1+\overline{w_X}(a))},
\end{equation}
where
\begin{equation}\label{eq:14}
\overline{w_X}(a)=\frac{\int w_X(a)d(\ln a)}{d(\ln a)}.
\end{equation}
The solution of nonhomogeneous equation (\ref{eq:12}) is
\begin{eqnarray}\label{eq:15}
\rho_X= & - & \left[ \int_1^a a^{n-1+3\overline{w_X}(a)}da\right] a^{-3(1+\overline{w_X}(a))} \nonumber\\
& + & \frac{C_X}{a^{3(1+\overline{w_X}(a))}}.
\end{eqnarray}
Finally we obtain
\begin{eqnarray}\label{eq:16}
\rho_{\text{eff}} & \equiv & 3H^2 +3\frac{k}{a^2}=\rho_m+\rho_{X} \nonumber \\
& = & \frac{C_m}{a^3}+\frac{\alpha}{n}a^{n-3} +\frac{C_X}{a^{3(1+\overline{w_X}(a))}} \nonumber \\
& - & \left[ \int_1^a a^{n-1+3\overline{w_X}(a)}da\right] a^{-3(1+\overline{w_X}(a))}.
\end{eqnarray}
The second and last terms origin from the interaction between dark matter and dark energy sectors.

Let us consider the simplest case of $\overline{w_X}(a)=$const$=w_X(a)$. Then integration of (\ref{eq:15}) can be performed and we obtain
\begin{equation}\label{eq:17}
\rho_{\text{eff}}=\frac{C_m}{a^3}+\frac{C_X}{a^{3(1+w_X)}}+\frac{C_\text{int}}{a^{3-n}}
\end{equation}
where $C_{\text{int}}=\frac{\alpha}{n}-\frac{\alpha}{n-3w_X}$. In this case we obtain one additional term in $\rho_{\text{eff}}$ or in the Friedmann first integral scaling like $a^{2-n}$. It is convenient to rewrite the Friedmann first integral to the new form using dimensionless density parameters. Then we obtain
\begin{eqnarray}\label{eq:18}
\left(\frac{H}{H_0}\right)^2 & = & \Omega_{\text{m},0}(1+z)^3+\Omega_{k,0}(1+z)^2 \nonumber \\
& + & \Omega_{\text{int}}(1+z)^{3-n} + \Omega_{X,0}(1+z)^{3(1+w_X)}.
\end{eqnarray}

Note that this additional power law term related to interaction can be also interpreted as the Cardassian or polytropic term \cite{Freese:2002sq,Godlowski:2003pd} (one can easily show that the assumed form of interaction always generates a correction of type $a^m, m=1-n$, in the
potential of the $\Lambda$CDM model and vice versa). Another interpretation of this term can origin from the Lambda decaying cosmology when the Lambda term is parametrized by the scale factor \cite{Costa:2007sq}.

In the next section we draw a comparison between the above model with the assumption that $\overline{w_X}(a)=\text{const}=-1$ and the $\Lambda$CDM model.

\section{Data}

To estimate the parameters of the both models we used the modified for our purposes \textsc{CosmoMC} code \cite{CosmoMC,Lewis:2002ah} with the implemented nested sampling algorithm \textsc{multinest} \cite{Feroz:2007kg,Feroz:2008xx}.

We used the observational data of 580 supernovae type Ia (the \textsc{Union2.1} compilation \cite{Suzuki:2011hu}), 31 observational data points of Hubble function from \cite{Jimenez:2001gg,Simon:2004tf,Gaztanaga:2008xz,Stern:2009ep,Moresco:2012jh,Busca:2012bu,Zhang:2012mp,Blake:2012pj,Chuang:2012qt,Anderson:2013oza} collected in \cite{Chen:2013vea}, the measurements of BAO (barion acoustic oscillations) from the Sloan Digital Sky Survey (SDSS-III) combined with the 2dF Galaxy Redshift Survey (2dFGRS) \cite{Eisenstein:2005su,Percival:2009xn,Eisenstein:2011sa,Ahn:2013gms}, the 6dF Galaxy Survey (6dFGS) \cite{Jones:2009yz,Beutler:2011hx}, the WiggleZ Dark Energy Survey \cite{Drinkwater:2009sd,Blake:2011en,Blake:2011wn}. We also used information coming from determinations of the Hubble function using the Alcock-Paczy\'{n}ski test \cite{Alcock:1979mp,Blake:2011ep}. This test is very restrictive in the context of modified gravity models.

The likelihood function for the supernovae type Ia data is defined by
\begin{equation}
L_{\text{SN}} \propto \exp \left[ - \sum_{i,j}(\mu_{i}^{\text{obs}} - \mu_{i}^{\text{th}}) C_{ij}^{-1} (\mu_{j}^{\text{obs}} - \mu_{j}^{\mathrm{th}})\right] , \label{sn_likelihood}
\end{equation}
where $C_{ij}$ is the covariance matrix with the systematic errors, $\mu_{i}^{\text{obs}}=m_{i}-M$ is the distance modulus, $\mu_{i}^{\text{th}}=5\log_{10}D_{Li} + \mathcal{M}=5\log_{10}d_{Li} + 25$, $\mathcal{M}=-5\log_{10}H_{0}+25$ and $D_{Li}=H_{0}d_{Li}$, where $d_{Li}$ is the luminosity distance which is given by $d_{Li}=(1+z_{i})c\int_{0}^{z_{i}} \frac{dz'}{H(z')}$ (with the assumption $k=0$).

For $H(z)$ the likelihood function is given by
\begin{equation}
L_{H_z} \propto \exp \left[ - \sum_i\frac{\left(H^{\text{th}}(z_i)-H^{\text{obs}}_i\right)^2}{2 \sigma_i^2} \right ],
\label{hz_likelihood}
\end{equation}
where $H^{\text{th}}(z_i)$ denotes the theoretically estimated Hubble function, $H^{\text{obs}}_i$ is observational data.

The likelihood function for the BAO data is characterized by
\begin{equation}
L_{\text{BAO}} \propto \exp \left[ - \sum_{i,j}\left(d^{\text{th}}(z_i)-d^{\text{obs}}_i\right) C_{ij}^{-1} \left(d^{\text{th}}(z_j)-d^{\text{obs}}_j\right)\right] \label{bao_likelihood}
\end{equation}
where $C_{ij}$ is the covariance matrix with the systematic errors, $d^{\text{th}}(z_i)\equiv r_s(z_d) \left[(1+z_i)^2 D_A^2(z_i)\frac{cz_i}{H(z_i)} \right ]^{-\frac{1}{3}}$, $r_s(z_d)$ is the sound horizon at the drag epoch and $D_A$ is the angular diameter distance.

The likelihood function for the information coming from the Alcock--Paczy\'{n}ski test is given by
\begin{equation}
L_{AP} \propto \exp \left[ - \sum_i\frac{\left(AP^{\text{th}}(z_i)-AP^{\text{obs}}_i\right)^2}{2 \sigma_i^2} \right]
\label{ap_likelihood}
\end{equation}
where $AP^{\text{th}}(z_i)\equiv \frac{H(z_i)}{H_0 (1+z_i)}$.

Finally, we used likelihood function for the CMB shift parameter $R$ \cite{Bond:1997wr}, which is defined by
\begin{equation}
L_{\text{CMB}} \propto \exp \left[ -\frac{1}{2}\frac{(R^{\text{th}}-R^{\text{obs}})^2}{\sigma_{\mathcal{A}}^2} \right]
\label{cmbr_likelihood}
\end{equation}
where $R^{\text{th}}=\frac{\sqrt{\Omega_{\text{m}} H_0}}{c}(1+z_{*})D_\mathcal{A}(z_{*})$, $D_\mathcal{A} (z_{*})$ is the angular diameter distance to the last scattering surface, $R^{\text{obs}}=1.7477$ and $\sigma_{\mathcal{A}}^{-2}=48976.33$ \cite{Li:2013awa}.

The total likelihood function $L_{\text{tot}}$ is defined as
\begin{equation}
L_{\text{tot}}=L_{\text{SN}}L_{H_z}L_{\text{BAO}}L_{\text{CMB}}L_{AP}.
\label{total_likelihood}
\end{equation}

\section{Results}

\subsection{The model parameter estimation}

The results of the estimation of parameters of the $\Lambda$CDM and the interacting $\Lambda$CDM models are presented in Table~\ref{tab1_values}.
Given the likelihood function (\ref{sn_likelihood}), first, we estimated the models parameters using the Union2.1 data only. Next, the parameter estimations with the joint data of the Union2.1, $h(z)$, BAO, Alcock-Paczy\'{n}ski test (likelihood functions (\ref{sn_likelihood}), (\ref{hz_likelihood}), (\ref{bao_likelihood}) and (\ref{ap_likelihood})) have been performed. At last, we estimated the model parameters with the joint data enlarged with CMB data (the total likelihood function (\ref{total_likelihood})).

The values of the interaction parameter $\Omega_{\text{int},0}$ is very small for all data sets. Especially the result for the second data set (Union2.1, $h(z)$, BAO, AP data) indicates that the interaction is probably negligible. There is also no indication of the direction of interaction if it is a physical effect. While for the Union 2.1 data set only the interaction parameter $\Omega_{\text{int},0}$ is negative and a greater value of $\Omega_{\text{m},0}$ in the interacting $\Lambda$ CDM model implies the flow from the dark energy sector to the matter sector, for the data set consisting of all data the opposite.

\begin{table*}
\caption{The mean of marginalized posterior PDF with $68 \%$ confidence level for the parameters of the models. In the brackets are shown parameter's values of joined posterior probabilities. Estimations were made using the Union2.1, $h(z)$, BAO, determinations of Hubble function using Alcock--Paczy\'{n}ski test and CMB R data sets.}
\label{tab1_values}
\begin{tabular*}{\textwidth}{@{\extracolsep{\fill}}cccccc@{}}
  \hline
  & Union2.1 data & Union2.1, $h(z)$, BAO, & Union2.1, $h(z)$, BAO, \\
  & only & AP data & AP, CMB data \\
  \hline
  \multicolumn{4}{c}{interacting model}\\
  \hline
    $\Omega_{\text{m},0} \in \langle 0,1 \rangle$    & $0.3126^{+0.0064}_{-0.0343} (0.2952)$ & $0.2770^{+0.0119}_{-0.0130} (0.2690)$ & $0.2847^{+0.0107}_{-0.0115} (0.2725)$ \\
    $\Omega_{\text{int},0} \in \langle -1,1 \rangle$ & $-0.0232^{+0.1070}_{-0.1018} (-0.3492)$ & $0.0109^{+0.0146}_{-0.0267} (0.0734)$ & $-0.0139^{+0.0244}_{-0.0056} (-0.0152)$ \\
    $m \in \langle -10,10 \rangle$            &$-0.2687^{+1.2726}_{-0.3223} (-0.0528)$ & $0.5622^{+0.7790}_{-0.5499} (0.9911)$ & $0.3205^{+0.7826}_{-0.6730} (3.7364)$ \\
    $h \in \langle 0.6,0.8 \rangle$           & $0.7004^{+0.0996}_{-0.1004} (0.7912)$ & $0.6949^{+0.0121}_{-0.0148} (0.6937)$ & $0.6957^{+0.0120}_{-0.0147} (0.7093)$ \\
  \hline
  \multicolumn{4}{c}{$\Lambda$CDM model}\\
  \hline
    $\Omega_{\text{m},0} \in \langle 0,1 \rangle$    & $0.2956^{+0.0035}_{-0.0034} (0.2955)$ & $0.2777^{+0.0070}_{-0.0073} (0.2791)$ & $0.2912^{+0.0043}_{-0.0045} (0.2904)$ \\
    $h \in \langle 0.6,0.8 \rangle$           & $0.7000^{+0.1000}_{-0.1000} (0.6053)$ & $0.6932^{+0.0048}_{-0.0049} (0.6922)$ & $0.6858^{+0.0041}_{-0.0043} (0.6849)$ \\
  \hline
\end{tabular*}
\end{table*}

\begin{figure}
\centering
   \includegraphics[width=0.35\textwidth]{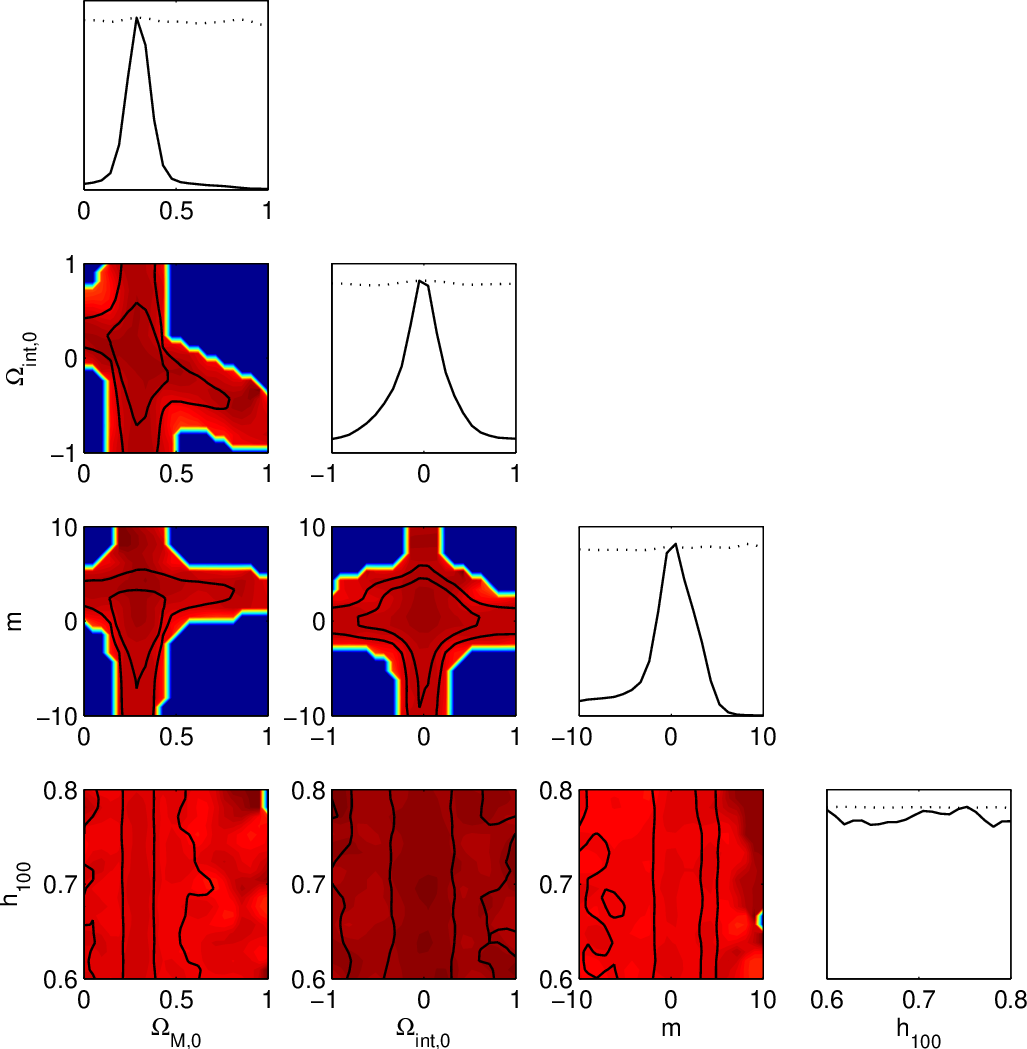}\\
   \includegraphics[width=0.35\textwidth]{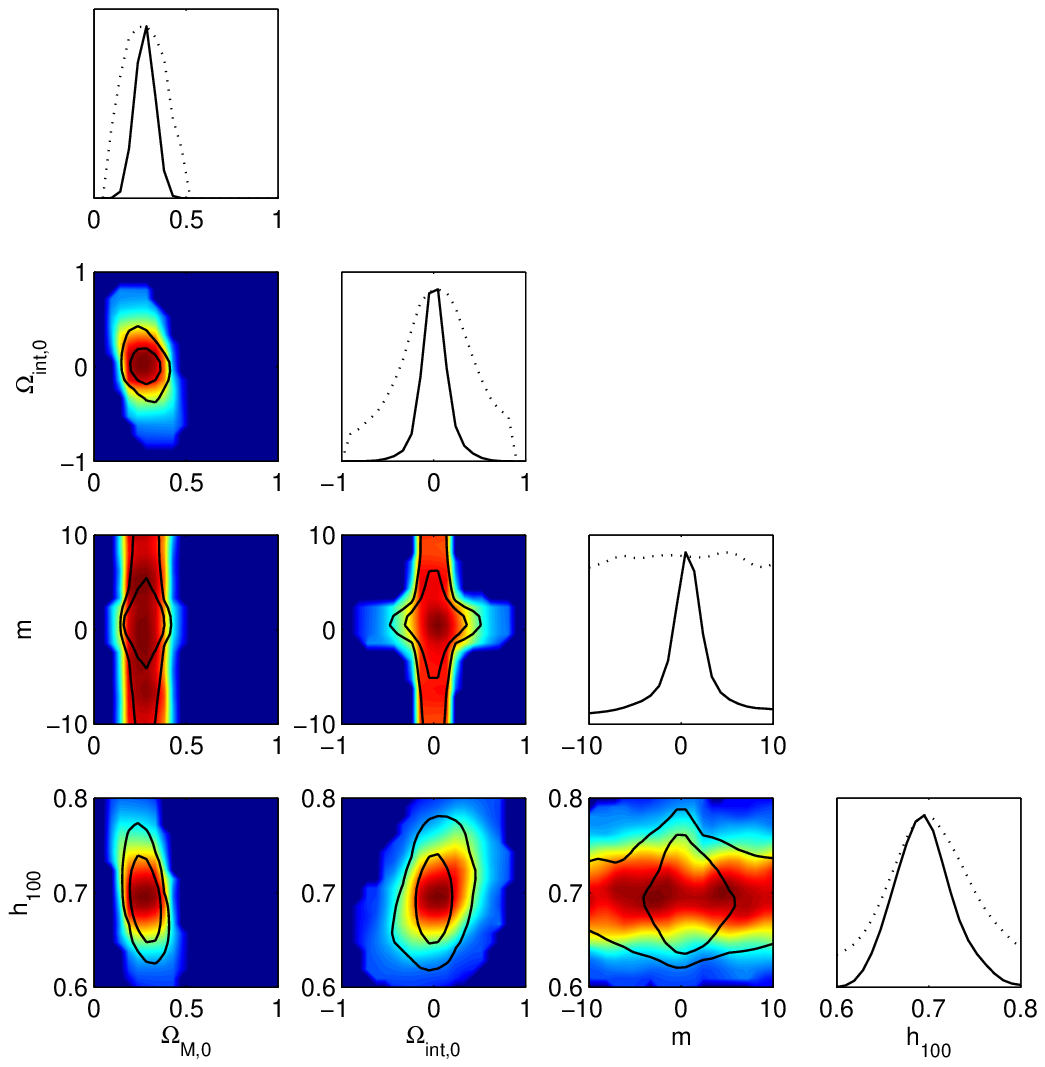}\\
   \includegraphics[width=0.35\textwidth]{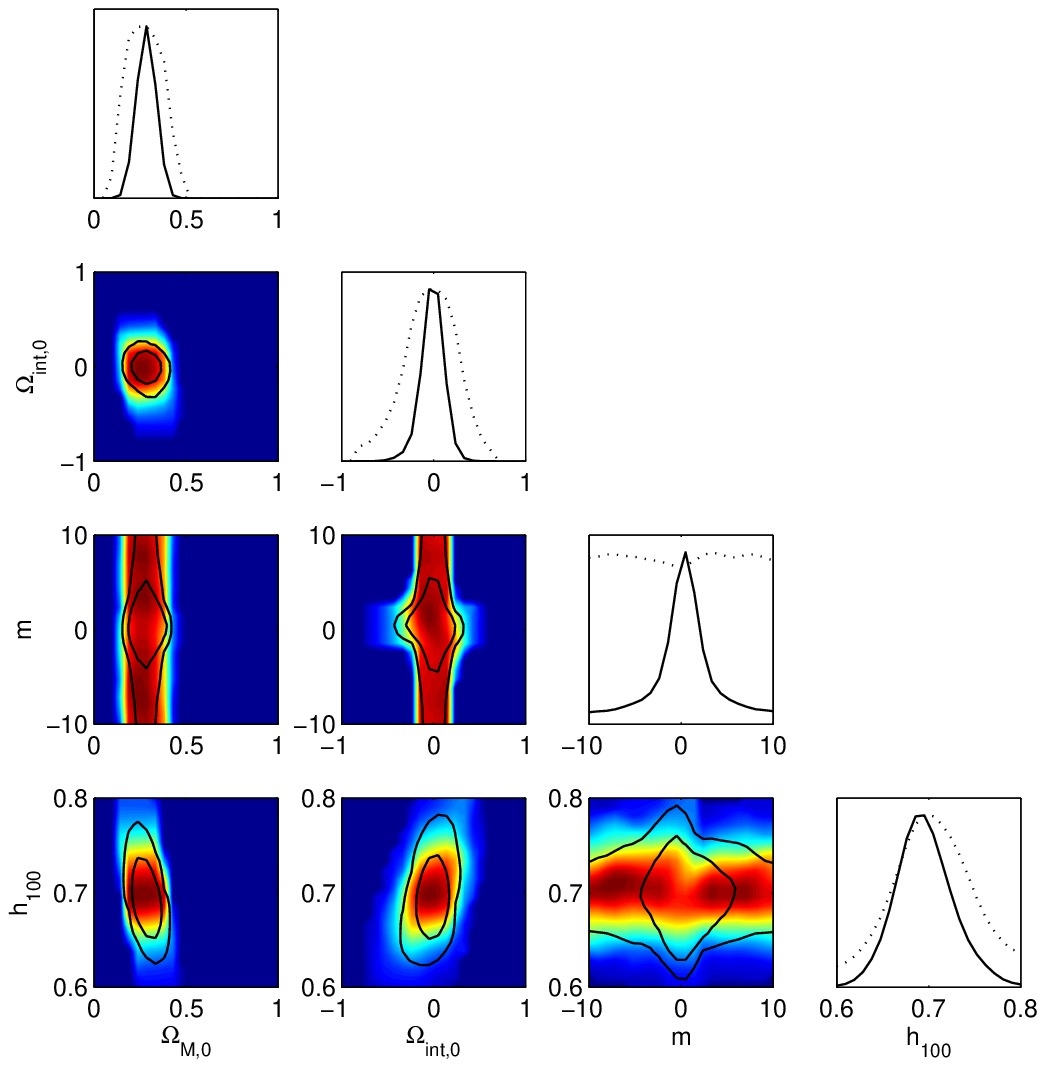}\\
  \caption{Posterior constraints for the interacting model. Joint probability distributions for $h_{100}$, $\Omega_{\text{M},0}$, $\Omega_{\text{int}}$ and $m$ with each other as well as marginalized probability distributions for each variable. Solid lines denote $68 \%$ and $95 \%$ credible intervals of fully marginalized probabilities, the colors illustrate mean likelihood of the sample. Top: estimations with the Union2.1 data only. Middle: estimations made using the Union2.1, $h(z)$, BAO, and determinations of the Hubble function using Alcock--Paczy\'{n}ski test data sets. Bottom: estimations made using the Union2.1, $h(z)$, BAO, determinations of the Hubble function using Alcock--Paczy\'{n}ski test and CMB R data sets.}
   \label{pos_mod_int_de}
\end{figure}

\begin{figure}
\centering
   \includegraphics[width=0.35\textwidth]{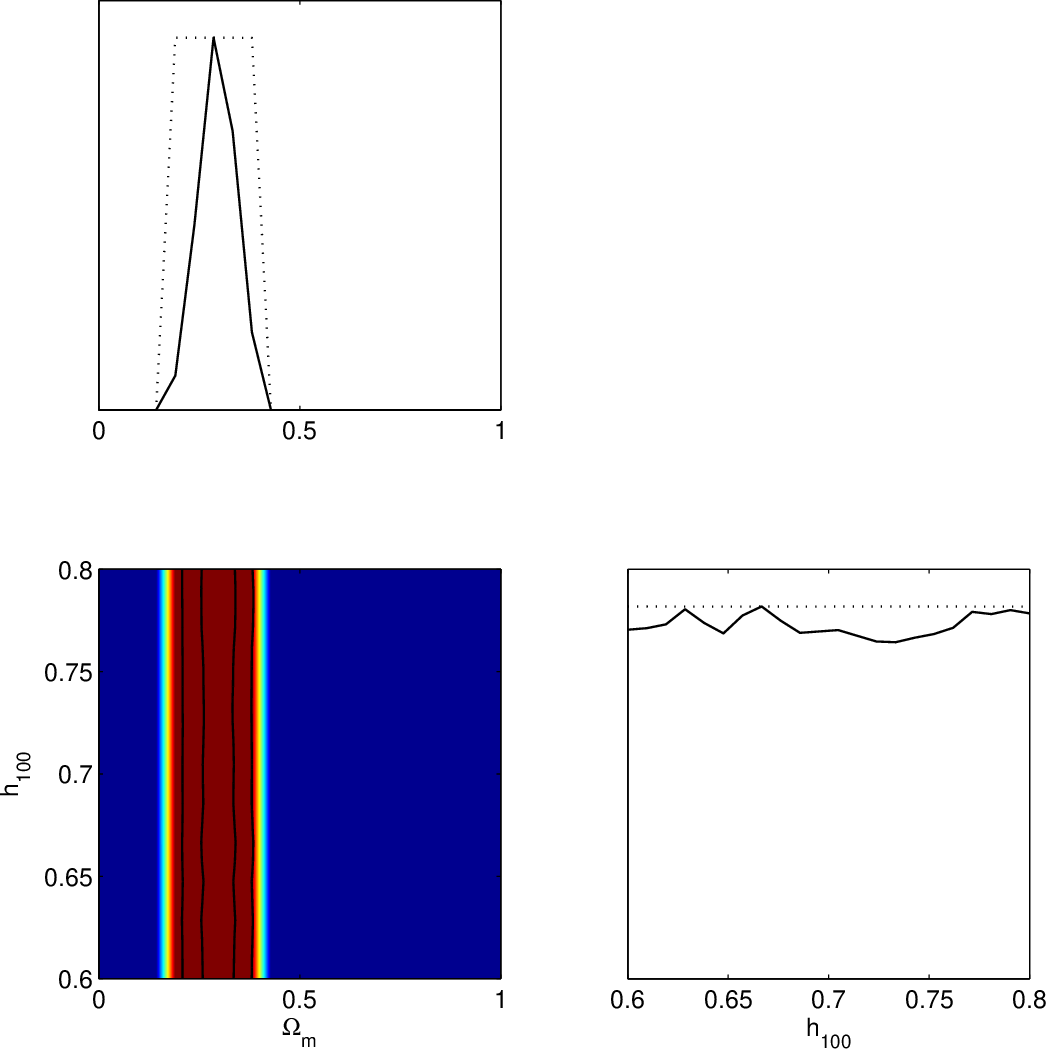}\\
   \includegraphics[width=0.35\textwidth]{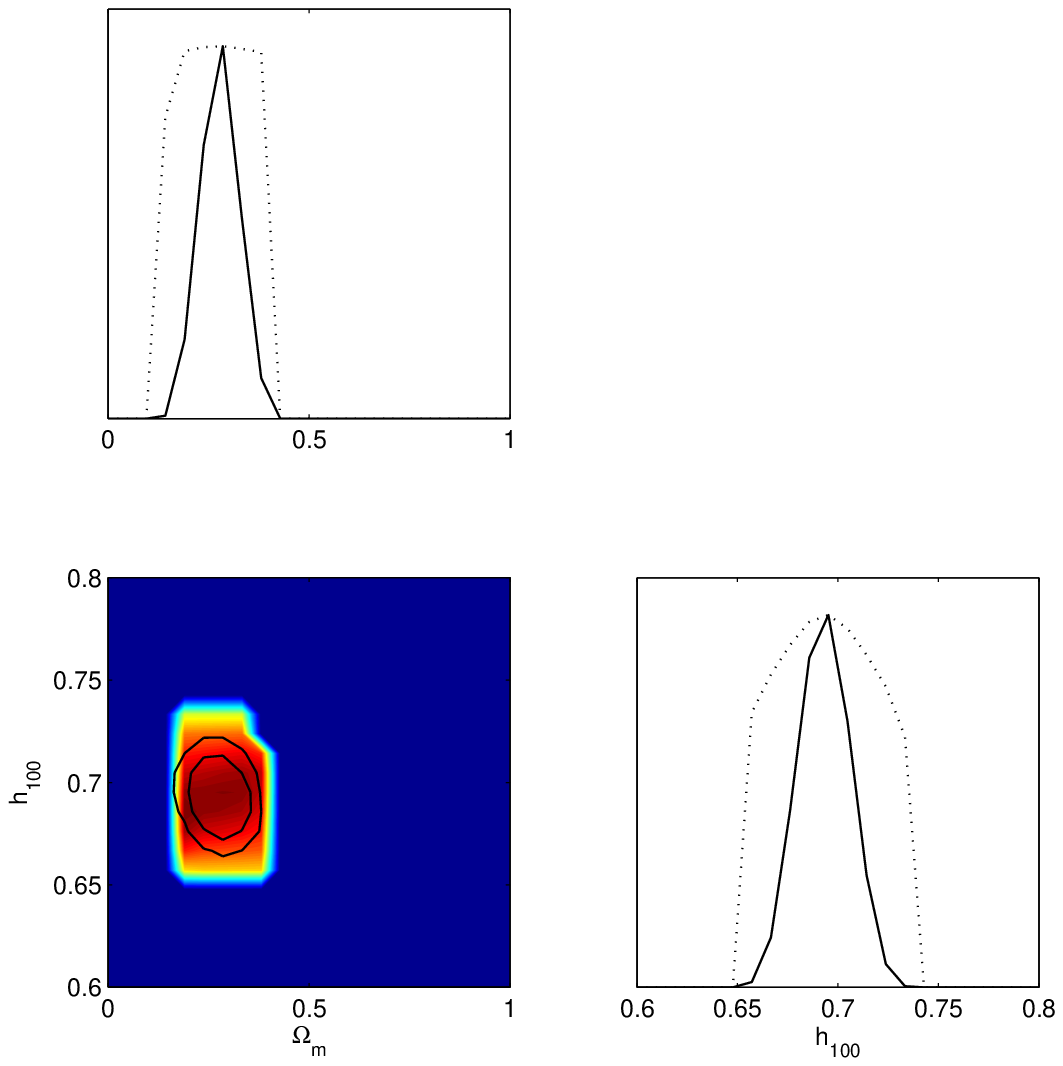}\\
   \includegraphics[width=0.35\textwidth]{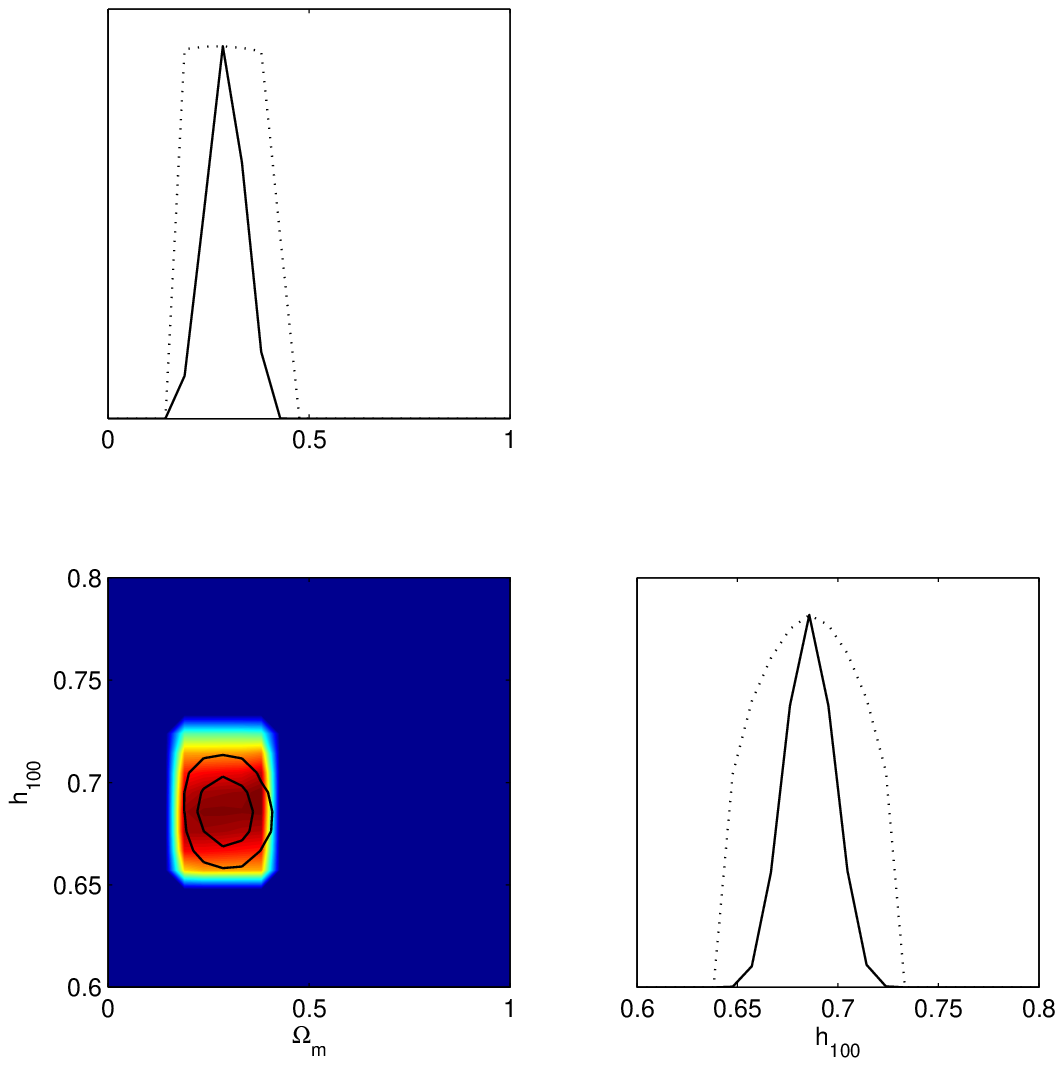}\\
   \caption{Posterior constraints for the $\Lambda$CDM model. Joint probability distributions for $h_{100}$, $\Omega_{\text{M},0}$ with each other as well as marginalized probability distributions for each variable. Solid lines denote $68 \%$ and $95 \%$ credible intervals of fully marginalized probabilities, the colors illustrate mean likelihood of the sample. Top: estimations with the Union2.1 data only. Middle: estimations made using the Union2.1, $h(z)$, BAO, and determinations of the Hubble function using Alcock--Paczy\'{n}ski test data sets. Bottom: estimations made using the Union2.1, $h(z)$, BAO, determinations of the Hubble function using Alcock--Paczy\'{n}ski test and CMB R data sets.}
   \label{pos_mod_lcdm}
\end{figure}

The uncertainty of the each estimated model parameter is presented twofold: as $68\%$ confidence levels in Table~\ref{tab1_values} and as the marginalized probability distributions in Fig.~\ref{pos_mod_int_de} and~\ref{pos_mod_lcdm}.

\subsection{The likelihood ratio test}

We begin our statistical analysis from the likelihood ratio test. In this test one of the models (null model) is nested in a second model (alternative model) by fixing one of the second model parameters. In our case the null model is the $\Lambda$CDM model, the alternative model is the interactive $\Lambda$CDM model, and the parameter in question is $\Omega_{\text{int}}$.
\begin{align*}
H_0 &\colon \Omega_{\text{int}}=0 \\
H_1 &\colon \Omega_{\text{int}} \neq 0.
\end{align*}
The statistic is given by
\begin{equation}\label{lik_ratio_test}
\lambda = 2 \ln \left( \frac{L(H_1|D)}{L(H_0|D)} \right) = 2\left( \frac{\chi^2_{\text{int}}}{2} - \frac{\chi^2_{\Lambda\text{CDM}}}{2} \right)
\end{equation}
where $L(H_1|D)$ is the likelihood of the interacting $\Lambda$CDM model, $L(H_0|D)$ is the likelihood of the $\Lambda$CDM model obtained with using three different sets of data. The statistic $\lambda$ has the $\chi^2$ distribution with $df=n_1-n_0=2$ degree of freedom where $n_1$ is number of the parameters of the alternative model, $n_0$ is number of the parameters of the null model. The results are presented in Table~\ref{classic_test}.
In all three cases the p-values are greater than the significance level $\alpha = 0.05$, that why the null hypothesis cannot be rejected. In other words we cannot reject the hypothesis that there is no interaction between the dark matter and dark energy sector.

\begin{table*}
\caption{The results of the likelihood ratio test for the $\Lambda$CDM model (null model) and the $\Lambda$CDM interacting model (alternative model). The values of $\chi^2_{\text{int}}$, $\chi^2_{\Lambda\text{CDM}}$, test statistic $\lambda$ and corresponding p-values ($df=4-2=2$). Estimations were made using the Union2.1, $h(z)$, BAO, determinations of the Hubble function using Alcock--Paczy\'{n}ski test, and CMB R data sets.}
\label{classic_test}
\begin{tabular*}{\textwidth}{@{\extracolsep{\fill}}lcccc@{}}
  \hline
data sets &$\chi^2_{\text{int}}/2$ & $\chi^2_{\Lambda\text{CDM}}/2$ & $\lambda$ & p-value \\
  \hline
Union2.1 & $272.5377$ & $272.5552$ & $0.0350$ & $0.9826$ \\
Union2.1, $h(z)$, BAO, AP & $282.2215$ & $282.2555$ & $0.0680$ & $0.9667$ \\
Union2.1, $h(z)$, BAO, AP, CMB & $282.3073$ & $282.4912$ & $0.3678$ & $0.8320$ \\
  \hline
\end{tabular*}
\end{table*}

\subsection{The model comparison using the AIC, BIC and Bayes evidence}

To obtain the values of AIC and BIC quantities we perform the $\chi^2=-2\ln L$ minimization procedure after marginalization over the $H_0$ parameter in the range $\langle 60,80 \rangle$. They are presented in Table~\ref{aic_bic}.

Regardless the data set the differences of the AIC quantities are in the interval $(3.4, 4)$, and are a little outside the interval $(4,7)$ which indicates the considerably less support for the interacting $\Lambda$CDM model. It means that while the $\Lambda$CDM model should be preferred over the interacting $\Lambda$CDM model, the latter cannot be ruled out.

However we can arrive at the decisive conclusion employing the Bayes factor. The difference of BIC quantities is greater than 10 and have the values in interval $(12,13)$ for all data sets. Thus, the Bayes factor indicates the strong evidence against the interacting $\Lambda$CDM model comparing to the $\Lambda$CDM model. Therefore we are strongly convinced to reject the interaction between dark energy and dark matter sectors due to Occam's principle.

\begin{table*}
\caption{Values of the $\chi^2$, AIC, $\Delta$AIC (with respect to the $\Lambda$CDM model), BIC and Bayes factor. Estimations were made using the Union2.1, $h(z)$, BAO, determinations of Hubble function using the Alcock--Paczy\'{n}ski test, and CMB R data sets.}
\label{aic_bic}
\begin{tabular*}{\textwidth}{@{\extracolsep{\fill}}lccccc@{}}
  \hline
data sets &$\chi^2/2$ & AIC & $\Delta \text{AIC}_{\text{int},\Lambda\text{CDM}}$ & BIC & $2\ln B_{\Lambda\text{CDM,int}}$ \\
  \hline
  \multicolumn{6}{c}{interacting model}\\
  \hline
Union2.1 & $272.5377$ & $553.0754$ & $3.9650$ & $570.5275$ & $12.6910$ \\
Union2.1, $h(z)$, BAO, AP & $282.2215$ & $572.4430$ & $3.9320$ & $590.1683$ & $12.7947$ \\
Union2.1, $h(z)$, BAO, AP, CMB & $282.3073$ & $572.6146$ & $3.6322$ & $590.3464$ & $12.4981$ \\
  \hline
  \multicolumn{6}{c}{$\Lambda$CDM model}\\
  \hline
Union2.1 & $272.5552$ & $549.1104$ & --- & $557.8365$ & --- \\
Union2.1, $h(z)$, BAO, AP & $282.2555$ & $568.5110$ & --- & $577.3736$ & --- \\
Union2.1, $h(z)$, BAO, AP, CMB & $282.4912$ & $568.9824$ & --- & $577.8483$ & --- \\
  \hline
\end{tabular*}
\end{table*}

\section{Conclusion}

We considered the cosmological model with dark energy represented by the cosmological constant and the model with the interaction between dark matter and dark energy (the interacting $\Lambda$CDM model). These models were studied statistically using the available astronomical data and then compared using the tools taken from the information as well as Bayesian theory. In both cases the model selection is based on Occam's principle which states that if two models describe the observations equally well we should choose the simpler one. According to the Akaike and Bayesian information criteria the model complexity is interpreted in the term of a number of free model parameters, while according to the Bayesian evidence a more complex model has a greater volume of the parameter space.

Anyone using the Bayesian methods in astronomy and cosmology should be aware of the ongoing debate not only about pros but also cons of this approach. Efstathiou provided a critique of the evidence ratio approach indicating difficulties in defining models and priors \cite{Efstathiou:2008ed}. Jenkins and Peacock called attention to too much noise in data which not allows to decide to accept or reject a model based solely on whether the evidence ratio reaches some threshold value \cite{Jenkins:2011va}. That is a reason that we also used the Akaike information criterion based on information theory.

The observational constraints on the parameter values, which we have obtained, have confirmed previous results that if the interaction between dark energy and matter is a real effect it should be very small. Therefore it seems to be natural to ask whether cosmology with interaction between dark energy and matter is plausible.

At the beginning of our model selection analysis we performed the standard likelihood ratio test. This test conclusion was to fail to reject the null hypothesis that there is no interaction between matter and dark energy sectors with the significance level $\alpha=0.05$. It was the first clue against the interacting $\Lambda$CDM model. The $\Delta$AIC between both models was less conclusive. While the $\Lambda$CDM model was more supported, the interacting $\Lambda$CDM cannot be rejected. On the other hand the Bayes factor have given decisive result, there was a very strong evidence against the interacting $\Lambda$CDM model comparing to the $\Lambda$CDM model. Given the weak or almost none support for the interacting $\Lambda$CDM model and bearing in mind Occam's razor we are inclined to reject this model.

We have also the theoretical argument against the interacting $\Lambda$CDM model. If we consider the $H^2$ formula which is a base for estimation there is a degeneracy because one cannot distinguish effects of interaction from the effect $w(z)$--the case of varying equation of state depending on time or redshift.

As was noted by Kunz \cite{Kunz:2007rk} there is the dark degeneracy problem. It means that the effect of interaction cannot be distinguished from the effect of an additional non-interacting fluid with the constant equation of state $w_{\text{int}}=n/3-2$. Therefore if we considered a mixture of all three non-interacting fluids we obtained the coefficient equation of state for the dark energy and interacting fluid in the form
\begin{align}
w_{\text{dark}} &= \frac{(p_X+ p_{\text{int}})}{C_X(1+z)^{3(1+w_X)} + C_{\text{int}}(1+z)^{3-n}} \nonumber \\
&= \frac{w_{X}(1+z)^{3(1+w_{X})} + C_{\text{int}}(1+z)^{3-n}w_{\text{int}}}{C_X(1+z)^{3(1+w_X)} + C_{\text{int}}(1+z)^{3-n}}.
\end{align}

\begin{acknowledgement}M. Szyd{\l}owski has been supported by the National Science Centre (Narodowe Centrum Nauki) grant 2013/09/B/ST2/03455.
M. Kamionka has been supported by the National Science Centre (Narodowe Centrum Nauki) grant PRELUDIUM 2012/05/N/ST9/03857. We thank the referee for carefully going through our manuscript.
\end{acknowledgement}

%\bibliography{astrophysics,astrophysmath}
%\bibliographystyle{spphys}

\end{document}